# Large Number, Dark Matter, Dark Energy, and the Superstructures in the Universe (with Extension)

Wuliang Huang
*Institute of High Energy Physics, Chinese Academy of Sciences*
*P.O. Box 918(3), Beijing 100049, China*

Xiaodong Huang
*Department of Mathematics, University of California, Los Angeles*
*Los Angeles, CA 90095, U.S.A.*

**Abstract:** Since there are dark matter particles $\nu$ with mass $\sim 10^{-1} eV$ in the universe, the superstructures with a scale of $10^{19}$ solar mass (large number $A \sim 10^{19}$) appeared around the era of the hydrogen recombination. The redshift z distributions of quasars support the existence of superstructures. Since there are superstructures in the universe, it is not necessary for the hypothesis of dark energy. While neutrino $\nu$ is related to electro-weak field, the fourth stable elementary particle $\delta$ ($m_\delta \sim 10^0 - 10^1 eV$) is related to gravitation-"strong" field, which suggests $p + \bar{p} \to n/\bar{n} + \bar{\delta}/\delta$ and that some new meta-stable baryons appeared near the $TeV$ region. Therefore, a twofold standard model diagram is proposed, and related to many experiment phenomena: the new meta-stable baryons' decays produce $\delta$ particles, which are helpful to explain the Dijet asymmetry phenomena at LHC of CERN, the different results for the Fermilab's data peak, etc; However, according to the (B-L) invariance, the sterile "neutrino" from "the event excess in the MiniBooNe" can not be the fourth neutrino but rather the $\delta$ particle; We think that the $\delta$ particles are related to the phenomenon about neutrinos FTL, and that anti-neutrinos are faster than neutrinos. FTL is also related to the cosmic inflation, singular point disappearance, and abnormal red shift of SN Ia. Some experiments and observations are suggested.
In the Extension section, we clarify "mass tree", our finite universe, cosmic dual expansions, dual SM etc. And the LHC can look for new particles with decay products graviton/$\delta$-particle and new interaction indeed.

Since the observation data of WMAP (Wilkinson Microwave Anisotropy Probe) was announced in 2003[1], cosmology entered a new era. From these data, the distribution of dark energy ($\Omega_\Lambda$), dark matter ($\Omega_d$), radiation ($\Omega_r$), and ordinary baryon matter ($\Omega_b$) in the universe was deduced by approximate calculation[2],[3], and a deeper understanding of inflation, elementary particles, Hubble constant, etc. was reached. In this paper, we are continuing the discussions from the previous papers[4],[5],[6] on dark matter, stable particle, meta-stable particle and the superstructure in the universe[7]–[11].

## Dark energy and the superstructure in the universe

The era around the hydrogen recombination is a special era in the evolution of the universe. In this era, the universe was becoming transparent and NBDMP (non-baryon dark matter particles - d particles) became non-relativistic particles[4],[5],[7]. The mass of d particle is: $m_d \sim A^{-0.5} m_p \sim 10^{-1} eV$ [4],[5], $A = \sqrt{\hbar c / G m_p^2}$. Where $A$ is the large number in the universe (see Ref [12], [13], [14], [4], [5]). During the non-relativization process, the superstructures with a mass scale of $M_F \sim \frac{m_{pl}^3}{m_d^2} \sim 10^{19}$ solar mass emerged[4] and broke away from cosmic expansion. The length scale of the superstructures from the observations at the present time is $r_F \sim 10^3 Mpc$ [11]. Such superstructures are feeble structures[7]. When they broke away from the cosmic expansion, the inner environment of the superstructure is not like a static universe, nor a standard Friedmann universe, but something in between[7],[15]. Perhaps, the inner environment of the superstructure can be described using an equivalent $\Lambda CDM$ model as in approximate calculation, then we have $\Lambda \neq 0$ and no dark energy existed in the superstructures.

Since the length scale of superstructure at the present time is $r_F \sim 10^3 Mpc$, we need to measure CMB angular power spectrum around $10^1$ degrees across the sky. Another important way is to measure the space distribution of quasars (see Appendix I) in more details.

## Neutrino, $\delta$ particle, and the ultra-high energy primary cosmic ray spectrum

It is well known that dark matter particle must be a stable particle. There are three types of stable particles in the universe: electron $e$ (electromagnetic interaction), neutrino $\nu$ (weak interaction), and proton $p$ (strong interaction). All of these particles are fermions. It is rational to postulate that there exists a fourth stable elementary particle, which is also a fermion with mainly "gravitational" interaction. This can be named the $\delta$ particle. $\nu$ and $\delta$ are the candidates of dark matter.

Dark matter exists everywhere in the universe, and also surrounds the earth (see normalization method for celestial bodies with two constituents in Ref [16]). The ultra-high energy primary cosmic ray spectrum (UEPCRS) measured on the surface of the earth can also reflect the interaction between primary cosmic rays and dark matter particles. As previously discussed [4],[6], the "knee" of the UEPCRS can be explained by the interaction between the primary cosmic ray with energy $E_p \sim 10^{15} eV$ and the neutrinos with mass $m_\nu \sim 10^{-1} eV$ ($p + \bar{\nu} \to n + \bar{e}$). Since there may exist the $\delta$ particles, the "ankle" of the UEPCRS can be explained by the interaction between primary cosmic ray with energy $E_p \sim 10^{18} eV$ and the $\delta$ particles with mass $m_\delta \sim 10^0 eV$ [17] ($p + \delta/\bar{\delta} \to n/\bar{n} + p$).

## A gravity-"strong" unification model

As mentioned above, ($p + \bar{\nu} \to n + \bar{e}$) hints electro-weak unification; and ($p + \delta/\bar{\delta} \to n/\bar{n} + p$) hints gravity-"strong" unification. This can also be enlightened from the scale table for "radius scale", "mass scale", "CBHR" (classical black hole radius), and "FSS" (free stream scale):

| radius scale | mass scale | Classical Black Hole Radius (CBHR) | Free Stream Scale (FSS) |
|---|---|---|---|
| Planck radius $r_{pl}$ | Planck mass $m_{pl} = A\, m_p$ | "Planck" CBHR $r_{pl} \approx G\, m_{pl}/c^2$ | |
| Proton radius $r_p = A\, r_{pl}$ | Proton mass $m_p = \tilde{A}^2 m_p$ | "strong" CBHR $r_p \approx \tilde{G} m_p / c^2$ | |
| compact star radius $r_{star} = A^2 r_{pl}$ | star mass $M_{star} = A^3 m_p$ | star CBHR $r_{star} \approx G\, M_{star}/c^2$ | proton FSS $M_{star} \approx m_{pl}^3 / m_p^2$ |
| compact superstructure radius $r_F = A^3 r_{pl}$ | superstructure mass $M_F = A^4 m_p$ | superstructure CBHR $r_F \approx G\, M_F / c^2$ | dark matter FSS $M_F \approx m_{pl}^3 / m_d^2$ |

*In the table (slightly altered from the scale table in Ref [4], [5]) $r_{pl} = \sqrt{G\hbar/c^3}$, $m_{pl} = \sqrt{\hbar c/G}$, $A = \sqrt{\hbar c/G m_p^2} = m_{pl}/m_p = M_F/M_{star} \approx 10^{19}$, $\tilde{A}^2 = A^2/(A^2)=1$, and $\tilde{G} = (A^2)G$.*

From this table, it is obvious that the proton radius is connected with a "strong" gravitation constant $\tilde{G}$. This means that $G$ "becomes" $\tilde{G}$ in microcosm scale, and there are close relationships between gravitation interaction and proton as that between electromagnetic interaction and electron. While there are stable particle $e$ ($m_e \approx 0.5$ MeV), meta-stable particle $\mu$ ($m_\mu \approx 106$ MeV) and $\tau$ ($m_\tau \approx 1.8$ GeV) in electro-weak field, there may exist new meta-stable baryons with mass near TeV in gravitation-"strong" field: perhaps $p'$ and $p''$, which are related to $\delta'$ and $\delta''$ as $p$ related to $\delta$. All of these can be summarized as a twofold standard model diagram in Appendix II.

## Discussions for the twofold standard model diagram

(1) There are two ways to obtain $\delta$ particles (included $\delta'$ and $\delta''$) at LHC of CERN. The first way is $p + \bar{p} \to n/\bar{n} + \bar{\delta}/\delta$, but the energy threshold is too high (corresponding to "ankle"), so the event is very rare. Another way is the decay of $p'$ or $p''$ ($p' \to p\bar{\delta}\delta'$ ...), which can be produced in $pp$, $p\bar{p}$, p-ion, ion-ion collisions. The decays produce $\delta$ particles, which are helpful to explain the Dijet asymmetry phenomena[18] at LHC of CERN, especially if the jets are abundant with protons.

(2) According to the Twofold Standard Model Diagram, the Fermilab's data peak[19] is related to meta-stable particles $p'$ (or $\pi'$). The different results from D0 team and CDF team at Fermilab are related to the different effects of the new particle $\delta$ (or $G$) in different detectors.

(3) The "Event Excess in the MiniBooNE Search for $\bar{\nu}_\mu \to \bar{\nu}_e$ Oscillations" (Ref [20]) supports the existence of sterile "neutrino". Since $\bar{\nu}_\mu$ has B-L=1, it is possible that the sterile "neutrino" is not the fourth (and the fifth) neutrino, but the first (and the second) $\delta$ particle. Thus, the B-L number is still invariable. This hints that the two parts of the twofold standard model are related. To check this possibility, we need to do SBL experiments above ground.

(4) As for the "neutrinos faster than light[21]", it means that the gravitational signatures' propagation is faster than light and the $\delta$ particles' propagation is also faster than light. Because of neutrino oscillation, some time the neutrinos become $\delta$ particles during their propagation and appear an average speed faster than light. If $\nu_\mu$ particles are faster than light (underground), we think that $\bar{\nu}_\mu$ may propagate faster than $\nu_\mu$ (underground). There are more discussions about FTL in Appendix III.

(5) The twofold standard model diagram suggests a new interaction (relax interaction), which is mentioned above as "strong" interaction. The gravitational interaction would be at first unified with the relax interaction as electromagnetic interaction is at first unified with weak interaction.

## Some Experiments and Observations

(i) The dark energy in fact is a superstructure effect. Since our university is a multi-constituents universe, the stable constituents' ($p, e, \nu, \delta$ ...) decoupling and becoming non-relativistic should leave some imprints in the data of CMB Anisotropy. Since superstructures are feeble structures[7], we need to measure CMB angular power spectrum around $10^1$ degrees across the sky, as well as the space distribution of quasars[22], in more details.

(ii) The "knee" of the UEPCRS ($p + \bar{\nu} \to n + \bar{e}, n \to p\bar{\nu}e$) suggests an excess of cosmic ray electrons and positrons at energies $\gtrsim 10^0 TeV$ [23]. If there are $\delta$ particles, an excess at energies $\sim 10^{2-3} TeV$ is related to the "ankle" of UEPCRS:
$p + \delta/\bar{\delta} \to n/\bar{n} + p, n/\bar{n} \to p\bar{\nu}e/\bar{p}\nu e$.

(iii) When accretion disks are consumed in quasars, it is possible that some point sources

with "mono-color" gravitational wave ($10^{14-15} Hz$) / infrared emission ($10^{13-14} Hz$) can be observed across the sky. This can help us explore dark matter particles' mass and the energy source mechanism in quasars[22].

# Extension

For the above sections (ArXiv: 0804.2080v8, Dec 2011), we make extensions as follows:

(1) We extend the scale table above to a "mass tree" as in Ref [4]v9, Oct 2015:

$$\begin{array}{cc} m_0 = M_0 & \\ m_1 & M_1 \\ m_2 & M_2 \\ m_3 & M_3 \\ m_4 & M_4 \\ m_5 & M_5 \\ m_6 & M_6 \\ m_7 & M_7 \\ m_8 & M_8 \end{array}$$

On the left side are the mass scales of stable particles in micro-cosmos:

$m_n = A^{-\frac{n}{2}} m_{pl}$, n = 0, 1, 2 ...     (with length scale $r_n$)

On the right side are the corresponding mass scales of celestial bodies in macro-cosmos:

$M_n = A^n m_{pl}$, n = 0, 1, 2 …     (with length scale $R_n$ and density $\rho_n$)

In which,     Planck mass $m_{pl} \sim \sqrt{\dfrac{\hbar c}{G}} \sim 10^{19} GeV$

Planck length $r_{pl} \sim \sqrt{\dfrac{\hbar G}{c^3}} \sim \dfrac{G m_{pl}}{c^2}$

then     Large Number $A \sim \dfrac{m_{pl}}{m_p} \sim \dfrac{r_p}{r_{pl}} \sim 10^{19}$

(2) We insert a mass scales ($m' \leftrightarrow M'$) between ($m_2 \leftrightarrow M_2$) and ($m_3 \leftrightarrow M_3$):

Suppose $m' = m_e$ (electron mass), then $M' \sim 10^6 M_2$ (i.e. $10^6 solar-mass$) that is Globular cluster/AGN mass scale.

(3) We insert another mass scales ($m'' \leftrightarrow M''$) between ($m_2 \leftrightarrow M_2$) and ($m_3 \leftrightarrow M_3$):

Suppose $M''$=Galaxy cluster mass scale, then $m'' \sim 10^1 eV$ that is $m_\delta$ mass scale.

(4) Now the mass tree has a result as follows:

| n | $m_n$ | $M_n$ | $R_n$ |
|---|---|---|---|
| 0 | $m_{pl}$ | $m_{pl}$ | |
| 1 | $m_1$ | $M_{LBH}$ | |
| 2 | $m_2 \sim m_p$ | $M_{star} \sim A^3 m_p$ | |
|   | $m' \sim m_e$ | $M' \sim M_{AGN}$ | |
|   | $m'' \sim m_\delta$ | $M'' \sim M_{galaxy-cluster}$ | |
| 3 | $m_3 \sim m_\nu$ | $M_{superstructure}$ | |
| 4 | $m_4$ | $M_4$ | $R_4 \sim A \cdot r_p$ |
| 5 | $m_5$ | $M_5 \sim A^3 M_{star}$ | $R_5 \sim r_p$ |
| 6 | $m_6$ | $M_6$ | $R_6 \sim r_p$ |
| 7 | $m_7$ | $M_7$ | $R_7 \sim r_p$ |
| 8 | $m_8 \sim m_A$ | $M_8 \sim A^3 \cdot A^3 M_{star}$ | $R_8 \sim r_p$ |

In this table there are stable particles $p, e, \delta, \nu$, where $p$ and $e$ are "bright matter"; $\delta$ and $\nu$ are dark matter. Also, there are five types of super light particles (background particles) $m_8, m_7, m_6, m_5, m_4$ and two types of super heavy particles $m_1, m_{pl}$.

(5) From $n=8$ to $n=4$, those represent the evolution of the universe at early era. Since $R_8 = R_7 = R_6 = R_5 = r_p$, our universe is finite and the earth is not at the center[4]. When the amount of observation data is large enough, the similar celestial phenomena are anisotropy, including anisotropy spectrum of CMB itself. The peaks of CMB anisotropy spectrum may reflect the mass spectrum of $\nu$ and $\delta$ particles.

(6) The inflation occurs at $n=5$ and the cosmic expansion continues after $n=4$ ($R_4$ evolves to $R_{CMB}$, $R_{CMB}$ is the present horizon of our universe). The reasons of cosmic dual expansions are the confines of two different categories of quarks ($q$ and $q_l$). So the Dual Standard Model (Ref [17] v4, July 2011) is preferred:

$$
\begin{array}{cccc cccc}
u & c & t & \gamma & u_l & c_l & t_l & G \\
d & s & b & g & d_l & s_l & b_l & g_l \\
\nu_e & \nu_\mu & \nu_\tau & Z^0 & \delta & \delta' & \delta'' & Z' \\
e & \mu & \tau & W^\pm & p & p' & p'' & W'
\end{array}
$$

Where $u_l$ $c_l$ $t_l$ $d_l$ $s_l$ $b_l$ are lept-quarks, $g_l$ is lept-gluon, $G$ is graviton. $Z'$, $W'$ are the gauge bosons about a new type of interaction related to $\delta$ particles. The speed of photon $\gamma$ is $c$. The speed of graviton $G$ is $c'$. In this model the mass (energy) span is limited, so the LHC has the ability to look for new particles indeed, which decay products related to $\delta$-particle and graviton.

## Appendix I: Superstructure and the quasars' distribution

From the quasar number count as a function of redshift, $N(z)$, there are several peaks between $z=0$ and $z=5$ [24]. Strong peaks in $N(z)$ are at $z \approx 0.24, 1.2,$ and $1.8$. Using the results of the Fourier spectral analysis in that paper, we roughly adopted the redshift values of peaks as follows:

| i | 1 | 2 | 3 | 4 | 5 | 6 |
|---|---|---|---|---|---|---|
| $z_i$ | 0.24 | 0.5 | 0.8 | 1.2 | 1.8 | 3.0 |
| $d_i$ (Gpc) | 0.86 | 1.55 | 2.15 | 2.75 | 3.40 | 4.22 |
| $\Delta d_i$ (Gpc) | 0.86 | 0.69 | 0.60 | 0.60 | 0.65 | 0.82 |

where $d_i$ is the distance, and $\Delta d_i = d_i - d_{i-1}$, $d_0 = 0$. The average value of $\Delta d_i$ is 0.7 Gpc ($\tilde{\gtrsim} 10^3 Mpc$). This is the length scale of superstructure in the universe [4].

The redshift z distribution of SNs Ia [25],[26],[27] also supports the existence of superstructures, but the peak's period for SNs Ia is about two times of that for quasars.

## Appendix II: Twofold Standard Model Diagram

We suggest the twofold Standard Model Diagram as follows:

| $u$ | $c$ | $t$ | $\gamma$ | | $u'$ | $c'$ | $t'$ | $G$ |
|---|---|---|---|---|---|---|---|---|
| $d$ | $s$ | $b$ | $g$ | | $d'$ | $s'$ | $b'$ | $g'$ |
| $\nu_e$ | $\nu_\mu$ | $\nu_\tau$ | $Z^0$ | | $\delta$ | $\delta'$ | $\delta''$ | $Z'$ |
| $e$ | $\mu$ | $\tau$ | $W^\pm$ | | $p$ | $p'$ | $p''$ | $W'$ |

Where $G$ is graviton; Z', W' are the gauge bosons about a new type of interaction (relax interaction), which is related to $\delta$ particles as weak interaction is related to $\nu$ particles. $u'$ and $d'$ are related to a stable particle ($u'u'd'$) with mass of TeV or a stable particle (electron) with mass of ~MeV.

## Appendix III: FTL (Fast than Light)

(1) We assume there is a critical cosmic density $\rho_{cr}$ in the early universe. When cosmic density $\rho > \rho_{cr}$ during the evolution of the universe, the speed of light $c$ becomes FTL ($\bar{c}$). If $\bar{c}$ increases together with $\rho$ and $\rho/\bar{c}^2 = \rho_{cr}/c^2 =$ Constant, the evolution equation of the universe has a simple form: $\dot{R}^2 \propto R^2$. That means the universe is in an era of inflation.

(2) Using the afore-mentioned hypothesis to the collapse process of black hole, we get a minimum collapse radius $r_{min}$: $r_{min}^2 \propto \frac{1}{G}(\frac{c^2}{\rho_{cr}})$ for black hole, and the black hole does not become a singular point.

(3) There are some local weak FTL regions in the observable universe, where the abnormal red shift appeared. For example, SN Ia region. It means that the universe does not seem to undergo an accelerating expansion.

______________________________________________________________

Email address: huangwl39@yahoo.com, huangwl@ihep.ac.cn, xhuang@ucla.edu